



\font\titlefont = cmr10 scaled\magstep 4
\font\sectionfont = cmr10
\font\littlefont = cmr5
\font\eightrm = cmr8

\def\ss{\scriptstyle}
\def\sss{\scriptscriptstyle}

\newcount\tcflag
\tcflag = 0  

\ifnum\tcflag = 0 \magnification = 1200 \fi  

\global\baselineskip = 1.2\baselineskip
\global\parskip = 4pt plus 0.3pt
\global\abovedisplayskip = 18pt plus3pt minus9pt
\global\belowdisplayskip = 18pt plus3pt minus9pt
\global\abovedisplayshortskip = 6pt plus3pt
\global\belowdisplayshortskip = 6pt plus3pt

\def\barsoff{\overfullrule=0pt}


\def\endignore{}
\def\ignore #1\endignore{}

\newcount\dflag
\dflag = 0


\def\monthname{\ifcase\month
\or January \or February \or March \or April \or May \or June%
\or July \or August \or September \or October \or November %
\or December
\fi}

\newcount\dummy
\newcount\minute  
\newcount\hour
\newcount\localtime
\newcount\localday
\localtime = \time
\localday = \day

\def\advanceclock#1#2{ 
\dummy = #1
\multiply\dummy by 60
\advance\dummy by #2
\advance\localtime by \dummy
\ifnum\localtime > 1440 
\advance\localtime by -1440
\advance\localday by 1
\fi}

\def\settime{{\dummy = \localtime%
\divide\dummy by 60%
\hour = \dummy
\minute = \localtime%
\multiply\dummy by 60%
\advance\minute by -\dummy
\ifnum\minute < 10
\xdef\spacer{0} 
\else \xdef\spacer{}
\fi %
\ifnum\hour < 12
\xdef\ampm{a.m.} 
\else
\xdef\ampm{p.m.} 
\advance\hour by -12 %
\fi %
\ifnum\hour = 0 \hour = 12 \fi
\xdef\timestring{\number\hour : \spacer \number\minute%
\thinspace \ampm}}}



\def\endtitle{}
\def\title#1\endtitle{\vskip.5in\titlefont
\global\baselineskip = 2\baselineskip
#1\vskip.4in
\baselineskip = 0.5\baselineskip\rm}

\def\endauthors{}
\def\authors#1\endauthors{#1}

\def\endabstract{}
\def\abstract#1\endabstract{\vskip .3in%
\centerline{\sectionfont\bf Abstract}%
\vskip .1in
\noindent#1}

\def\nopageonenumber{\footline={\ifnum\pageno<2\hfil\else
\hss\tenrm\folio\hss\fi}}  

\newcount\nsection
\newcount\nsubsection

\def\section#1{\global\advance\nsection by 1
\nsubsection=0
\bigskip\noindent\centerline{\sectionfont \bf \number\nsection.\ #1}
\bigskip\rm\nobreak}

\def\subsection#1{\global\advance\nsubsection by 1
\bigskip\noindent\sectionfont \sl \number\nsection.\number\nsubsection)\
#1\bigskip\rm\nobreak}

\def\topic#1{{\medskip\noindent $\bullet$ \it #1:}}
\def\endtopic{\medskip}

\def\appendix#1#2{\bigskip\noindent%
\centerline{\sectionfont \bf Appendix #1.\ #2}
\bigskip\rm\nobreak}


\newcount\nref
\global\nref = 1

\def\therefs{}


\def\ref#1#2{\xdef #1{[\number\nref]}
\ifnum\nref = 1\global\xdef\therefs{\item{[\number\nref]} #2\ }
\else
\global\xdef\oldrefs{\therefs}
\global\xdef\therefs{\oldrefs\vskip.1in\item{[\number\nref]} #2\ }%
\fi%
\global\advance\nref by 1
}

\def\listrefs{\vfill\eject\section{References}\therefs}


\newcount\nfoot
\global\nfoot = 1

\def\foot#1#2{\xdef #1{(\number\nfoot)}
\footnote{${}^{\number\nfoot}$}{\eightrm #2}
\global\advance\nfoot by 1
}


\newcount\nfig
\global\nfig = 1
\def\thefigs{} 

\def\figure#1#2{\xdef #1{(\number\nfig)}
\ifnum\nfig = 1\global\xdef\thefigs{\item{(\number\nfig)} #2\ }
\else
\global\xdef\oldfigs{\thefigs}
\global\xdef\thefigs{\oldfigs\vskip.1in\item{(\number\nfig)} #2\ }%
\fi%
\global\advance\nfig by 1 } 

\def\fig#1{\xdef #1{(\number\nfig)}
\global\advance\nfig by 1 } 


\newcount\cflag
\newcount\nequation
\global\nequation = 1
\def\eqlabel{(1)}

\def\nexteqno{\ifnum\cflag = 0
\global\advance\nequation by 1
\fi
\global\cflag = 0
\xdef\eqlabel{(\number\nequation)}}

\def\lasteqno{\global\advance\nequation by -1
\xdef\eqlabel{(\number\nequation)}}

\def\label#1{\xdef #1{(\number\nequation)}
\ifnum\dflag = 1
{\escapechar = -1
\xdef\draftname{\littlefont\string#1}}
\fi}

\def\clabel#1#2{\xdef\eqlabel{(\number\nequation #2)}
\global\cflag = 1
\xdef #1{\eqlabel}
\ifnum\dflag = 1
{\escapechar = -1
\xdef\draftname{\string#1}}
\fi}

\def\cclabel#1#2{\xdef\eqlabel{#2)}
\global\cflag = 1
\xdef #1{\eqlabel}
\ifnum\dflag = 1
{\escapechar = -1
\xdef\draftname{\string#1}}
\fi}


\def\eeq{}

\def\eqnn #1\eeq{$$ #1 $$}

\def\eq #1\eeq{
\ifnum\dflag = 0
{\xdef\draftname{\ }}
\fi 
$$ #1
\eqno{\eqlabel \rlap{\ \draftname}} $$
\nexteqno}







\def\eqa #1\eeq{
\ifnum\dflag = 0
{\xdef\draftname{\ }}
\fi 
$$ \eqalignno{ #1 } $$
\global\cflag = 0}


\def\ie{{\it i.e.\/}}
\def\eg{{\it e.g.\/}}


\def\mpla#1#2#3{{\it Mod.\ Phys.\ Lett.} {\bf A#1}, (19#2) #3}

\def\npb#1#2#3{{\it Nucl.\ Phys.} {\bf B#1} (19#2) #3}
\def\plb#1#2#3{{\it Phys.\ Lett.} {\bf #1B} (19#2) #3}

\def\prd#1#2#3{{\it Phys.\ Rev.} {\bf D#1} (19#2) #3}
\def\pr#1#2#3{{\it Phys.\ Rev.} {\bf #1} (19#2) #3}

\def\prl#1#2#3{{\it Phys.\ Rev.\ Lett.} {\bf #1} (19#2) #3}


\global\nulldelimiterspace = 0pt



\def\frac#1#2{{{#1} \over {#2}}\,}  
\def\hf{{1\over 2}}
\def\nth#1{{1\over #1}}


\def\Dsl{\hbox{/\kern-.6700em\it D}} 
\def\dsl{\hbox{/\kern-.5300em$\partial$}}
\def\pxpsl{\hbox{/\kern-.5600em$p$}}
\def\ssl{\hbox{/\kern-.5300em$s$}}
\def\epssl{\hbox{/\kern-.5100em$\epsilon$}}
\def\delsl{\hbox{/\kern-.6300em$\nabla$}}
\def\lxpsl{\hbox{/\kern-.4300em$l$}}
\def\elxpsl{\hbox{/\kern-.4500em$\ell$}}
\def\kxpsl{\hbox{/\kern-.5100em$k$}}
\def\qxpsl{\hbox{/\kern-.5000em$q$}}
\def\sla#1{\raise.15ex\hbox{$/$}\kern-.57em #1}



\def\twi{\widetilde}

\def\roughly#1{\mathrel{\raise.3ex\hbox{$#1$
\kern-.75em\lower1ex\hbox{$\sim$}}}}

\def\tw#1{\tilde{#1}}





\def\Sca{{\cal A}}
\def\Scb{{\cal B}}

\def\Scl{{\cal L}}

\def\Scr{{\cal R}}


\def\ssa{{\sss A}}

\def\ssd{{\sss D}}
\def\sse{{\sss E}}

\def\ssg{{\sss G}}

\def\ssi{{\sss I}}

\def\ssl{{\sss L}}
\def\ssm{{\sss M}}


\def\det{\mathop{\rm det}}






\nopageonenumber
\baselineskip = 16pt
\barsoff


\def\abar{\overline{a}}

\def\ap{\alpha'}

\def\tt{\tilde{t}} 
\def\Fbar{\twi{F}}
\def\Gbar{\twi{G}}
\def\Hbar{\twi{H}}
\def\Kbar{\twi{K}}
\def\fbar{\twi{f}}
\def\gbar{\twi{g}}
\def\hbarr{\twi{h}}
\def\phibar{\twi{\phi}}
\def\Abar{\twi{A}}
\def\Bbar{\twi{B}}

\def\abar{\twi{\Sca}}
\def\bbar{\twi{\Scb}}
\def\Mbar{\twi{M}}
\def\Nbar{\twi{N}}
\def\Qbar{\twi{Q}}

\def\GN{G_{\scriptscriptstyle N}}
\def\bk{\item{}}
\def\IR{\relax{\rm I\kern-.18em R}}

\def\IR{\relax{\rm I\kern-.18em R}}
\font\cmss=cmss10 \font\cmsss=cmss10 at 7pt
\def\IZ{\relax\ifmmode\mathchoice
{\hbox{\cmss Z\kern-.4em Z}}{\hbox{\cmss Z\kern-.4em Z}}
{\lower.9pt\hbox{\cmsss Z\kern-.4em Z}}
{\lower1.2pt\hbox{\cmsss Z\kern-.4em Z}}\else{\cmss Z\kern-.4em Z}\fi}


\line{hep-th/9411195 \hfil McGill-94/41, NEIP-94-008}

\title
\centerline{Duality and}
\centerline{Four-Dimensional Black Holes}
\endtitle

\authors
\centerline{C.P. Burgess,${}^a$ R.C. Myers,${}^a$ and F. Quevedo${}^b$}
\vskip .15in
\centerline{\it ${}^a$ Physics Department, McGill University}
\centerline{\it 3600 University St., Montr\'eal, Qu\'ebec, Canada, H3A 2T8.}
\vskip .1in
\centerline{\it ${}^b$ Institut de Physique, Universit\'e de Neuch\^atel}
\centerline{\it CH-2000 Neuch\^atel, Switzerland.}
\endauthors

\vskip .1in
\abstract
We consider the effects of abelian duality transformations on static,
spherically-symmetric, asymptotically flat string spacetimes in four
dimensions, where the dilaton, axion, metric, and gauge fields
are allowed to be nonzero. Independent of the $\ap$ expansion, there
is a six-parameter family of such configurations, labelled by
the charges characterizing the asymptotic behaviour of the various
fields: \ie\ their mass, dilaton charge, axion charge, electric charge,
magnetic charge, and Taub-NUT parameter. We show that duality,
based on time-translation invariance, maps these solutions amongst
themselves, with the effect of interchanging two pairs of these six
labels, namely: (1) the mass and dilaton charge, and (2) the axion charge
and the Taub-NUT parameter.  We consider in detail the special case
of  the purely Schwarzschild black hole, for which the mass of the
dual configuration vanishes to leading order in $\ap$.
Working to next-to-leading order in $\ap$ for the bosonic and
heterotic strings, we find that duality takes a black hole of mass $M$
to a (singular) solution having mass $\sim - \, 1/(\ap M)$.
Finally, we argue that two solutions
which are related by duality based on a noncompact symmetry
are {\it not} always physically equivalent.
\endabstract


\vfill\eject

\section{Introduction}

\ref\duality{For a recent review see, \eg\ A. Giveon, M. Porrati and E.
Rabinovici, {\it Phys. Rept.} {\bf 244} (1994) {77};
E. Alvarez, L. Alvarez-Gaum\'e and Y. Lozano, ``An Introduction to
T-Duality in String Theory,'' preprint CERN-TH-7486, hep-th/9410237.}
\ref\buscher{T. Buscher, \plb{194}{87}{59}; \plb{201}{88}{466}.}

One of the remarkable features of string theory is the surfeit of symmetry
that it exhibits. This abundance of symmetry is also reflected in the
low-energy equations which govern the so-called massless modes of
the string: the dilaton, graviton, axion and gauge fields. Among the
most `stringy' of these symmetries are target space
duality transformations \duality,
which in their simplest form are discrete transformations that relate
physically equivalent string configurations. Such transformations are
guaranteed to exist whenever a spacetime admits a continuous
isometry \buscher.

\ref\largesmall{K. Kikkawa and Y. Yamasaki, \plb{149}{84}{357};\bk
N. Sakai and I. Senda, {\it Prog. Theor. Phys. Suppl.}
{\bf 75} (1986) 692.}
\ref\twodbh{E. Witten, \prd{44}{91}{314}}
\ref\moretwodbh{R. Dijkgraaf, E. Verlinde and
H. Verlinde, \npb{371}{92}{269};\bk A. Giveon \mpla{6}{91}{2843}.}
\ref\bpb{J. Horne and G. Horowitz, \npb{368}{92}{444};\bk
J. Horne, G. Horowitz and A. Steif, \prl{68}{92}{568}.}
\ref\paulandme{P. Ginsparg and F. Quevedo, \npb{385}{92}{527}.}
\ref\mirror{See for example {\it Essays on Mirror Manifolds},
edited by S.-T. Yau, International Press, Hong Kong (1992).}
\ref\sduality{A. Font, L.E. Ib\'a\~nez, D. L\"ust and F. Quevedo,
\plb{249}{90}{35};\bk
S.J. Rey, \prd{43}{91}{526}.}

Duality symmetries are particularly interesting because they
have been found to relate string solutions that one would
otherwise have thought to be completely unrelated. The original
such example is the duality between string propagation on very
large and very small toroidal spacetimes \largesmall, but there
are many others. Target space duality relates strongly-curved,
singular regions and ordinary nonsingular parts of spacetime,
such as arise in two-dimensional black holes \twodbh, \moretwodbh,
and in black string spacetimes \bpb, \paulandme. Other duality
symmetries include mirror symmetry \mirror,
and the conjectured duality between weak and strong coupling
in string theory \sduality.

\ref\simp{C.R. Nappi and E. Witten, \prl{71}{93}{3751};
S.B. Giddings, J. Polchinski and A. Strominger, \prd{48}{93}{5784};
S.B. Giddings and A. Strominger, \prl{67}{91}{2930};
R.C. Myers, \plb{199}{87}{371};
I. Antoniadis, C. Bachas, J. Ellis and D.V. Nanopoulos,
\plb{211}{88}{393}; \npb{328}{89}{117}; C.V. Johnson
\prd{50}{94}{4032}; E. Kiritsis, C. Kounnas and
D. L\"ust, \plb{331}{94}{321}.}
\ref\exten{G.T. Horowitz and A.A. Tseytlin, ``A new class of exact
solutions in string theory,'' preprint UCSBTH-94-31, hep-th/9409021;
``Extremal black holes as exact string solutions,'' preprint
IMPERIAL-TP-93-94-51, hep-th/9408040; ``On exact solutions and
singularities in string theory,'' IMPERIAL-TP-93-94-38, hep-th/9406067;
I. Antoniadis, S. Ferrara and C. Kounnas, \npb{421}{94}{343}.
}
\ref\fandx{X. de la Ossa and F. Quevedo, \npb{403}{93}{377}.}

To date, comparatively little has been done to explore the implications of
target space duality for spacetimes in higher --- most notably
four --- dimensions, largely due to lack of knowledge in this case of the
explicit form for the solutions to the string equations to all orders in $\ap$,
except in certain special cases \simp. (For some recent progress in
finding exact solutions see Ref.~\exten.)
Our purpose in the present paper is to partially fill this gap, by examining
some of the implications of duality in four (and higher) dimensions.
We therefore consider here the generic static, spherically-symmetric, and
asymptotically-flat field configurations consisting of the metric,
dilaton, axion and electromagnetic fields --- in fact below, we relax
the restriction that the solutions are static. Such a field configuration
always admits a duality transformation that is based on its invariance
under time translation (as well as other dualities \fandx\ based
on the rotational symmetries), and it is to this duality that we refer in
what follows.

\ref\bmq{C.P. Burgess, R.C. Myers and F. Quevedo, ``On Spherically
Symmetric String Solutions in Four Dimensions'', preprint McGill-94/47,
NEIP-94-011, hep-th/9410142.}

We are able to circumvent the obstacle of not knowing the exact solutions
to the string equations by working, as much as possible, with the generic field
configuration which is consistent with the assumed symmetries, and
with asymptotic flatness. We also couch our discussion of the action of
duality on these configurations
completely in terms of its implications for the charges of the solutions which
describe their asymptotic falloff at spatial infinity. This permits us to
describe
the exact action of duality in a way which does {\it not} rely on an explicit
knowledge of the functional form of the explicit solutions to all orders in
$\ap$.
When an explicit form for the fields does become necessary, we use the
lowest-order in $\ap$ solutions which have been recently found \bmq\
for the six-parameter family of configurations which we consider.

A general time-independent and spherically-symmetric field configuration
is completely characterized by a small number of functions of a
single (radial) variable, $r$.  Since the string field equations, to
lowest order in the $\ap$ expansion, are second order partial differential
equations, the $r$ dependence of their static solutions is typically
determined by the asymptotic behaviour of the various fields
as $r \to \infty$. For asymptotically-flat configurations involving the
metric, dilaton, axion and electromagnetic fields, we argue that this
asymptotic
behaviour is specified by the values of five parameters, or `charges'.
This counting of parameters is not affected by the inclusion of the
higher-order
corrections in $\ap$ --- despite the fact that these typically involve higher
derivatives --- since we consider only solutions which have a well-behaved
limit as $\ap$ tends to zero.

The five charges that arise in this way are the configuration's ($i$)
mass, ($ii$) dilaton charge, ($iii$) axion charge, ($iv$) electric charge and
($v$) magnetic charge. If we relax the condition that the metric be static,
and instead just require it to be stationary,\foot\differenc{We explain this
distinction later in the text.} then other parameters become
possible. One such is the metric's Taub-NUT parameter, which we include
since, as we shall see, it arises naturally in the action of duality on
solutions to the
low-energy string equations. This six-parameter family of field configurations
includes the string extensions of many familiar spacetimes, such as the
Schwarzschild and charged dilatonic black holes. Of these the
Schwarzschild black hole is probably the simplest since --- to lowest order in
$\ap$ --- all of its charges vanish, except for its mass.

We find that duality has a very simple action on this six-parameter family,
interchanging two pairs of the asymptotic constants.
In particular, it interchanges ($i$) the configuration's mass with its dilaton
charge, and ($ii$) the Taub-NUT parameter with the axion charge. By
contrast, the electric and magnetic charges remain invariant. Somewhat
surprisingly, we find that these conclusions sensitively depend on the
asymptotic behaviour that is assumed for the electromagnetic gauge
potential, $A_\mu$, which we take here (in four dimensions) to vanish
like $1/ r$.

\ref\strbhone{C.G. Callan, R.C. Myers and M.J. Perry, \npb{311}{89}{673}.}

Of particular interest is the duality transformation of the nonsingular
dilaton--metric black hole of mass $M$, for which the dilaton
charge happens to be zero
to leading order in $\ap$. In this case duality produces a solution
which is singular at the Schwarzschild radius, and which has a mass
which is also zero at this order. Taking into account the
$O(\ap)$ corrections \strbhone, however, we find that the
mass $M$ black hole is actually mapped to a solution whose mass is
negative, and given by $- k/(\ap M)$,
with $k$ a positive dimensionless constant for the bosonic
and heterotic strings. For the superstring $k=0$, and so it is
necessary to work to higher order in $\ap$ in order to determine
the mass of the dual solution. We find that this calculation gives
a dual mass for the superstring which is again negative, and of order
$- 1/(\ap^3 M^5)$ in size.

\ref\kiritsis{A. Giveon and E. Kiritsis, \npb{411}{94}{487}.}
\ref\compacteq{M. Ro\v cek and E. Verlinde, \npb{373}{92}{630}.}

This connection, under duality, between solutions with opposite signs for the
mass  is reminiscent of the duality between the asymptotically flat
region (positive mass) and the region beyond the singularity (negative
mass) of the 2D black hole, but unlike that case, the 4D black hole
seems to map two different geometries instead of two different regions
of the same geometry. This might be puzzling if duality transformations were
believed to relate
physically equivalent configurations, as has been argued for some
Wess-Zumino-Witten (WZW) models \kiritsis. Indeed, although
the physical equivalence of the dual solutions is expected to hold
whenever the symmetry on which the dualization is based is compact
\compacteq, their equivalence for noncompact symmetries is not yet clear.
We use our explicit dual solutions to argue here that duality could relate
physically inequivalent string solutions. One way in which we do
so is by demonstrating that they give different results
for the scattering, at  large radii, of massless string states.

The paper is organized as follows. In the next section we identify the
class of field configurations which satisfy our assumed symmetries
and boundary conditions. In so doing we identify the six
charges which define the asymptotic behaviour of the various fields.
We then, in section 3, derive the action of duality on these solutions.
Section 4 explores in detail the action of duality for the
special case of solutions which involve just the dilaton and metric fields.
We start by finding the action of duality on the solutions to the field
equations at lowest-order in $\ap$. We then focus on the black-hole
solution, \ie, that which contains a nonsingular horizon. Since the
mass of the dual solution, in this case,
actually vanishes at lowest order in $\ap$, we find the higher-order
corrections and obtain the mass of the dual solution to this order.
This involves a short digression concerning the
$O(\ap)$ corrections to the duality transformations themselves.
Section 5 addresses the physical (in)equivalence of the dual
configurations. We first compute the propagation of massless
string states by the general dilaton-metric configuration. We
then use the results of this calculation to argue that duality based
on noncompact symmetries may not relate physically equivalent
string backgrounds.  Our conclusions are finally summarized in section 6.
We discuss in an appendix the $d > 4$ case for the
metric--dilaton--electromagnetic gauge-field system.

\section{Static Spherically Symmetric String Solutions}

\ref\lestrings{See, \eg\ M. Green, J. Schwarz and E. Witten
{\it Superstring Theory I} (1987), (Cambridge University Press).}
\ref\calyau{P. Candelas, G.T. Horowitz, A. Strominger and E. Witten,
\npb{258}{85}{46}.}

The three fields which appear (in string perturbation theory) in
the massless spectrum of a generic string theory consist of the metric,
$G_{\mu \nu}$, the dilaton, $\phi$, and an antisymmetric Kalb-Ramond field,
$B_{\mu \nu}$. In the heterotic string these can also be supplemented by
one or more gauge potentials, $A_\mu$. Additional moduli are also
possible for specific string vacua. The low-energy action for these fields
can be written, in four spacetime dimensions, as \lestrings:
\label\leaction
\eq
\Scl = {1 \over 8 \pi\ap}\; \sqrt{- G}\; e^\phi
\left[ R(G) + (\nabla \phi)^2 - {1 \over 12} \, H^{\mu\nu\lambda}
H_{\mu\nu\lambda} - {1\over 8} \, F^{\mu\nu}F_{\mu\nu} \right] + \cdots \, ,
\eeq
where $H_{\mu\nu\lambda} =  \partial_\mu B_{\nu\lambda}
- \nth{4} \, A_\mu F_{\nu\lambda} + $ (cyclic permutations), and
$F_{\mu\nu} = \partial_\mu A_\nu - \partial_\nu A_\mu$ are the
appropriate field strengths.  $R(G)$ here denotes the Ricci scalar for the
`sigma-model' metric, $G_{\mu\nu}$, and $\sqrt{-G} = \sqrt{-\det
G_{\mu\nu}}$ denotes the usual volume element. The ellipses in
this equation represent
terms which involve other massless fields and/or terms involving more
derivatives that arise at higher orders in the $\ap$ expansion. Since
no cosmological constant is included here, we imagine that our solutions
are complemented by some conformal field theory, such as a toroidal or
Calabi--Yau \calyau\ compactification, whose central charge
ensures conformal invariance on the world-sheet.

For future use we record here the rescaling that takes one to the `Einstein
frame', for which the coefficient of the scalar curvature in
the action is independent of the dilaton, $\phi$:
\label\esmrelation
\eq  g_{\mu\nu} \equiv  e^\phi\,  G_{\mu\nu}.
\eeq

\vfill\eject
\subsection{Static Spherically Symmetric Configurations}

\ref\exactgrsolns{D. Kramer, H. Stephani, E. Herlt, M. MacCallum
and E. Schmutzer, {\it Exact Solutions of Einstein's Field Equations},
(1980) (Cambridge University Press).}
\ref\weinberg{See, \eg\ S.~Weinberg, {\it Gravitation and Cosmology:
Principles and Applications of the General Theory of Relativity} (1982),
(New York: Wiley); or C.W. Misner, K.S. Thorne and J.A. Wheeler,
 {\it Gravitation} (1973), (San Francisco: Freeman).}

In four spacetime dimensions, the most general static and spherically
symmetric metric may always be written \exactgrsolns, \weinberg
\label\sssmetric
\eq
dS^2 \equiv G_{\mu\nu} \, dx^\mu dx^\nu = - F(r) \, dt^2 + G(r) \, dr^2
+ H(r)(d\theta^2 + \sin^2\theta \, d\varphi^2) .
\eeq
In fact, the additional freedom in reparameterizing the radial
coordinate is commonly used to fix either $H(r)=r^2$ or
$H(r)=r^2G(r)$, but neither choice is particularly convenient
for our present analysis.
Similarly, the generic time-independent and spherically symmetric
configuration for the other fields is
\label\sssdilem
\eq
\phi = \phi(r), \qquad
F_{tr} = E(r), \qquad
F_{\theta \varphi} = B(r) \sin \theta.
\eeq
For the Kalb-Ramond field, the time-independent, spherically
symmetric configurations are most easily identified through
the scalar `axion' field which is dual to the antisymmetric
tensor field:
\label\axion
\eq
H_{\mu\nu\rho} = - e^{-\phi}\epsilon_{\mu\nu\rho\kappa}\nabla^\kappa a
\eeq
where $\epsilon$ is the volume form for the sigma-model metric.
Now the generic time-independent, spherically
symmetric configuration includes: $a = a(r)$.

Given these configurations, the corresponding Einstein metric is
\label\sssemetric
\eq
ds^2 \equiv g_{\mu\nu} \, dx^\mu dx^\nu = - f(r) \, dt^2 + g(r) \, dr^2 +
h(r) (d\theta^2 + \sin^2\theta \, d\varphi^2),
\eeq
where $f = F \, e^{\phi}$, $g = G \, e^{\phi}$ and $h = H \, e^{\phi}$.

\ref\moresolns{M.J. Bowick, S.B. Giddings, J.A. Harvey, G.T. Horowitz,
and A. Strominger, \prl{61}{88}{2823}.}

There are two important kinds of field configurations which we omit
with the above symmetry ansatz. The first of these is the possible
topological Kalb-Ramond configuration, $B_{\theta\phi} =  Q_{\rm top}
\sin\theta$, for which the corresponding curl vanishes: $dB = 0$. Such
a field configuration is not pure gauge provided that the second
homotopy group of the background spacetime is nontrivial \moresolns.

The second class of configurations which need not be captured by the
above ansatz are those spacetimes which are time-independent,
but are not static. That is to say, those for which it is impossible to choose
a time coordinate such that the `time-space' components of the
metric vanish everywhere. Such spacetimes are often called stationary.

\ref\taubnut{A.H. Taub, {\sl Ann. Math.} {\bf 53}
(1951) 472; E. Newman, L. Tamburino and T. Unti, {\sl J. Math. Phys.}
{\bf 4} (1963) 915.}
\ref\misner{C.W. Misner, `Taub-NUT space as a counterexample to almost
anything', in {\it Relativity Theory and Astrophysics I, Relativity and
Cosmology},
ed. by J. Ehlers, Lectures in Applied Mathematics, Vol. 8 (American
Mathematical
Society, 1967) 160.}

Since we shall find that duality transformations
need not preserve the static form for the metric, we wish to
broaden our metric ansatz to include some of these more general
configurations. In particular we choose to work with a stationary
line element which is reminiscent of the `Taub-NUT' metric \taubnut,
\misner:
\label\TNUTansatz
\eq
dS^2 \equiv G_{\mu\nu} \, dx^\mu dx^\nu = - F(r) \, (dt + 2 N \cos\theta
\, d\varphi)^2 + G(r) \, dr^2 + H(r) (d\theta^2 + \sin^2\theta
\, d\varphi^2) ,
\eeq
and similarly for the Einstein metric. The parameter $N$ which
appears here we call the Taub-NUT parameter of the metric. Clearly,
the static case corresponds to the choice $N=0$. The metric of
eq.~\TNUTansatz\ is still invariant under time translations and
$SO(3)$ `rotations' \misner, but the latter rotations also act on the time
coordinate in order to preserve the differential $dt + 2 N \cos\theta
\, d\varphi$. Thus spherical symmetry in the conventional sense is lost.
Further the time coordinate must be periodically identified with period
$8\pi N$ in order to avoid conical singularities at the axes
$\theta=0$ and $\theta=\pi$ \misner.

\subsection{Asymptotic Forms}

The requirement of asymptotic flatness imposes some conditions on
the limiting forms as $r\to \infty$ of the various functions
that appear in the above {\it ans\"atze} for the low energy fields.
In particular, in this limit we may choose the metric functions $F$ and $G$
to approach unity, and the function $H$ to approach $r^2$. The
remaining fields are also restricted since asymptotic flatness
requires the stress-energy tensor to approach zero at large radii.

Therefore we take the two scalar fields, $\phi$ and $a$,  to approach constants
for large $r$. Since the axion field, $a$, as defined by eq.~\axion,
appears in equations of motion only through its derivative, the theory
admits the symmetry $a \to a + c$, for constant $c$. It is convenient to use
this symmetry to ensure that $a$ approaches zero as $r \to \infty$.

Similarly, it is also convenient to absorb the asymptotic value, $\phi_0$,
of the dilaton field into the definition for Newton's constant, $\GN$.
For instance, comparing the action of eq.~\leaction\ with the standard
Einstein-Hilbert form gives: $\GN = \hf \, e^{-\phi_0} \, \ap$. With this
choice we may also take the  dilaton field to vanish at infinity, in which
case the large-$r$ limit of the functions $f$ and $g$ of the Einstein
metric is also unity.

Finally, we require the electric and magnetic fields to fall to
zero at infinity, and this is accomplished if the component $F_{tr}$
of the electromagnetic field strength falls to zero, while the component
$F_{\theta\varphi}$ approaches $Q_\ssm \sin\theta$, for $Q_\ssm$ a
constant.

Solving the low-energy string equations in four spacetime dimensions
and imposing these asymptotic flatness conditions, typically produces
fields which approach their asymptotic values like $1/r$.
Therefore the solutions exhibit the following asymptotic behaviour:
\label\Gasymform
\eq
F(r) = 1 - {A \over r} + \cdots; \qquad G(r) = 1 + {B \over r} + \cdots; \qquad
e^{\phi(r)} = 1 - {Q_\ssd \over r} + \cdots \, ,
\eeq
and $H(r) = r^2 \left[ 1 + O\!\left({1 \over r} \right) \right]$.

The above relations imply similar ones for the Einstein metric:
\label\gasymform
\eq
f(r) = 1 - {\Sca \over r} + \cdots; \qquad g(r) = 1 + {\Scb \over r} + \cdots ;
\qquad h(r) = r^2 \left[ 1 + O\!\left( {1 \over r} \right) \right] \, ,
\eeq
with $\Sca = A + Q_\ssd$ and $\Scb = B - Q_\ssd$.

The asymptotic form for the remaining fields is taken to be
\label\emaform
\eq
a(r) = - \; {Q_\ssa \over r} + \cdots; \qquad
F_{tr} = {Q_\sse \over r^2} + \cdots, \qquad
\hbox{and} \qquad
F_{\theta\varphi} = Q_\ssm\, \sin\theta + \cdots \, .
\eeq
The ellipsis in the above expressions indicate terms of higher order in $1/r$.
The constants $Q_\ssa$, $Q_\sse$ and $Q_\ssm$ represent the
configuration's total axion, electric and magnetic charges, respectively.
Similarly in eq. \Gasymform, $Q_\ssd$ is called the dilaton charge.
Since the duality transformations in the presence of electromagnetic
fields are given in terms of the gauge potential, $A_\mu$, it is sometimes
necessary to work directly with the asymptotic form for this field.
Explicitly, we take
\label\amuform
\eq
A_t = {Q_\sse \over r} + \cdots, \qquad
\hbox{and} \qquad
A_{\varphi} =-\; Q_\ssm\, \cos\theta + \cdots ,
\eeq
while the remaining components vanish. In particular notice that
we take $A_t$ to vanish as $r \to \infty$.
As we shall see, many results turn out to be
sensitive to a nonzero value for $A_\mu$ at large $r$.
We also require the asymptotic form of the antisymmetric
tensor field corresponding to $a(r)$ in eq.~\emaform,
\label\hbform
\eq
H_{t\theta\varphi} = - \; Q_\ssa \sin\theta + \cdots; \qquad
\hbox{and} \qquad
B_{\varphi t} = Q_\ssa\, \cos\theta + \cdots ,
\eeq
while again the remaining components vanish.

\ref\ADM{R. Arnowitt, S. Deser and C.W. Misner, in {\it Gravitation:
An Introduction to Current Research}, ed. by L. Witten (New York, Wiley,
1962).}

The asymptotic forms for the metric have a physical significance
in that they determine the mass of the corresponding configuration.
The spatial components of the Einstein metric determine the conserved
(ADM) energy \ADM, or `inertial mass', $M_\ssi$, according
to the formula
\label\inertialmass
\eq
M_\ssi = {\Scb \over 2\GN} .
\eeq
This is the mass which emerges in a calculation of the energy using a
gravitational stress-energy pseudo-tensor, \weinberg. The asymptotic form
of $g_{tt}$ also has a significance, in that it determines the
`gravitational mass', $M_\ssg$, which controls the Newtonian limit of
the motion of nonrelativistic particles along geodesics in the
asymptotically-flat region. It is given by
\label\gravitationalmass
\eq
M_\ssg = {\Sca \over 2\GN} \ \ .
\eeq

Notice we use the asymptotic form for the Einstein metric, $g_{\mu\nu}$,
in the above expressions. Since we know that as $\ap \to 0$
the equations of motion for $g_{\mu\nu}$ reduce to Einstein's equations
with a conventional stress-energy tensor, we are guaranteed that
$M_\ssi=M_\ssg$ for the string solutions.
The latter result would not be true in general if the same definitions were
applied to the sigma-model metric. In that case, terms linear in the dilaton
enter stress-energy tensor, and can lead to
a difference between $M_\ssi$ and $M_\ssg$ \strbhone.

Finally, we remind the reader that the Taub-NUT parameter is determined
by the asymptotic behavior: $G_{t\varphi}=-2N\cos\theta+\cdots$.

\section{Duality}

We next turn to the behaviour of these solutions under duality
transformations based on time-translation symmetry. In order to do
so we first describe the transformation rules themselves.

\subsection{The Transformation Rules}

The existence of a continuous symmetry in a solution to the low-energy
string equations immediately permits the construction of another, dual,
solution to the same equations \duality. If the symmetry on which
this duality construction is based should be compact, then the two
configurations that are
related in this way represent precisely the same conformal field theory, and
so are two representations of a single string background \compacteq.
If the symmetry is noncompact, however, then the physical equivalence
of the dual solutions is not so clear. One of our motivations for studying
duality for higher-dimensional black holes is to pursue this issue in a
concrete setting.

\ref\welch{D.L. Welch, preprint UCSBTH-94-15, hep-th/9405070}
\ref\AALb{E. Alvarez, L. Alvarez-Gaum\'e and Y. Lozano,
\npb{424}{94}{155}, hep-th/9406206.}
\ref\aabl{E. Alvarez, L. Alvarez-Gaum\'e, J. Barb\'on and Y. Lozano,
\npb{415}{94}{71}}

If all of the fields are independent of the time coordinate,
the action of the duality transformation
for a nontrivial configuration involving the metric,
dilaton and antisymmetric tensors is given by
\buscher:
\label\dualbusch
\eq\eqalign{
\twi{G}_{tt} &= 1/G_{tt},\qquad
         \twi{G}_{ti}=-B_{ti}/G_{tt},\qquad
          \twi{G}_{ij} = G_{ij} -
{G_{ti}G_{tj} - B_{ti} B_{tj}\over G_{tt}}\cr
\twi{B}_{ti}& = -{G_{ti}}/{G_{tt}},\qquad
        \twi{B}_{ij}=B_{ij}+{G_{ti}B_{tj}
         -G_{tj}B_{ti}\over G_{tt}},\qquad
e^{\twi\phi}=e^\phi\left({\det G\over \det\twi G}\right)^{1/2}}
\eeq
where `$t$' denotes the time direction.\foot\funnysign{Our
transformation of the `$\ss t-i$' components of the fields differs from
that of Ref.~\buscher\ by a sign. They are nevertheless equivalent since
this sign may be compensated by performing the coordinate
transformation $\ss t\to -t$ in the dual solution.
A similar result was found in Refs.~\welch\ and \AALb.
Note that if we had made the alternate choice for these signs, then
the sign of the electric charge would be reversed under a duality
transformation.}
Ref.~\aabl\ presents these transformations in a manifestly
covariant framework based on the Killing vector associated with the
time-translation symmetry.

\ref\grv{A. Giveon, E. Rabinovici and
G. Veneziano, \npb{322}{89}{167};\bk
A. Shapere and F. Wilczek, \npb{320}{89}{669}.}
\ref\senn{S. Hassan and A. Sen, \npb{375}{92}{103};\bk
A. Sen, \plb{271}{91}{295};\bk
J. Maharana and J.H.  Schwarz, \npb{390}{93}{3}.}

The duality transformation is more complicated than
eqs.~\dualbusch\ when gauge fields, $A_\mu$, are present,
however. In this case, eqs.~\dualbusch\ may still be used, but with
the proviso that $G_{\mu\nu}$ is to be replaced everywhere
 by the quantity $G_{\mu\nu}+{1\over 4}A_\mu A_\nu$.
The only exception to this substitution is in the dilaton transformation,
which remains exactly the same as in eq.~\dualbusch.
The gauge potentials in the dual solution are given by \grv:
\label\dualgaugx
\eq
\eqalign{
\twi{A}_t&=- \; {A_t\over G_{tt}+{1\over4}A_tA_t}\cr
\twi{A}_i&=A_i-A_t \; {G_{ti}+{1\over4}A_tA_i-B_{ti}\over
G_{tt}+{1\over4}A_tA_t}\ \ .\cr}
\eeq

Two uncertainties hang over the above transformation rules in the presence
of gauge fields. Firstly, reference \grv\ considers only toroidal
compactifications, and so a derivation of these transformation rules
for gauge fields in the general case, along the lines of Ref.~\buscher,
is still lacking. One way to convince oneself that
Ref.~\grv's transformation rules are nonetheless correct in the general case,
however, is to identify the duality transformation as a discrete automorphism
of a larger class of $O(d,d+p)$ symmetries of the heterotic string,
as is done in Ref.~\senn. An independent construction of these
gauge field transformations begins with a Kaluza-Klein compactification
of the $(D+1)$-dimensional theory, and then makes a consistent truncation
(see for example the first article in Ref.~\exten) leaving the
$D$-dimensional theory with one extra $U(1)$ gauge field.
The above transformations \dualgaugx\  then follow directly
from the standard duality rules \dualbusch\ of the higher dimensional
theory\foot\ark{We thank A. Tseytlin for this
observation.}.

\ref\bergshoeff{E. Bergshoeff, I. Entrop and R. Kallosh, \prd{49}{94}{6663};
E. Bergshoeff, talk given at the 7th Marcel Grossman Meeting on General
Relativity, Stanford, July 24-30, 1994; E. Bergshoeff, T. Ort\'{\i}n
and R. Kallosh, in preparation.}

The second potential difficulty arises because other workers \bergshoeff\
have arrived at a different set of transformation rules in the presence of
gauge fields (following Ref.~\buscher\ ). We believe the origin of the nominal
contradiction between the results of Ref.~\bergshoeff\ on one hand, and
Refs.~\grv\ and \senn\ on the other lies in their different treatment of the
size of the gauge potential in the $\ap$ expansion. In one approach, gauge
fields
contribute at higher orders in the $\ap$ expansion of the low
energy string equations. This feature is important in determining the
supersymmetry of solutions, such as for Calabi-Yau compactifications.
Thus in Ref.~\bergshoeff, where the focus is on supersymmetry,
it appears that background gauge fields produce $\ap$ corrections
to the leading order duality transformations given in eqs.~\dualbusch.
In some circumstances, such as when studying black hole configurations,
one considers very large electromagnetic charges of $O(1/\sqrt{\ap})$,
which then contribute in the leading order equations. In this case, it is
convenient to rescale $A_\mu$ so that the gauge fields appear amongst
the field equations at leading order in $\ap$.
In the latter approach, we believe the results of Refs.~\grv\ and \senn\ apply.

\subsection{Asymptotic Forms in Four Dimensions}

The above duality transformations have the property that they
preserve the large-$r$ behaviour of the fields we are considering.
In particular, the form of the line element given in eq.~\TNUTansatz\
is preserved. So duality takes any configuration
from the family of solutions whose asymptotic forms are labelled by
the parameters, $A$, $B$, (or $\Sca$ and $\Scb$, for the Einstein metric),
$Q_\ssd$, $Q_\ssa$, $Q_\sse$, $Q_\ssm$ and $N$, onto another field
configuration that is also labelled by (usually different) values of the
same parameters. Hence the effect
of the duality transformation is conveniently expressed through
its action on the physical charges characterizing the
asymptotic fields.

If the parameters which define the dual solution
are denoted by $\twi A$, $\twi B$ (or $\twi \Sca$ and $\twi \Scb$
for the Einstein metric) $\twi Q_\ssd$, $\twi Q_\ssa$, $\twi Q_\sse$,
$\twi Q_\ssm$ and $\twi N$, then eqs.~\dualbusch\ and \dualgaugx\
imply these are given by:
\label\smasymdual
\eq
\twi A = - A, \qquad \twi B = B, \qquad \twi Q_\ssd = Q_\ssd + A, \qquad
\twi Q_\ssa = 2N, \qquad 2 \twi N = Q_\ssa ,
\eeq
while $\twi Q_\sse = Q_\sse$ and $\twi Q_\ssm = Q_\ssm$.

In terms of the constants $\Sca$ and $\Scb$ which give the
asymptotic behaviour of the dual Einstein metric the above equations become
\label\easymdual
\eq
\twi \Sca =  Q_\ssd, \qquad \twi \Scb = \Scb - \Sca + Q_\ssd, \qquad \twi
Q_\ssd =  \Sca .
\eeq

These last relations can be re-expressed, using eqs.~\inertialmass\ and
\gravitationalmass, in terms of the masses and dilaton charge.
We then see quite generally that, for an {\it arbitrary} member
of our six-parameter family of field configurations, duality has the following
action upon the asymptotic charges:
\label\dualsum
\eq
2 \GN M \leftrightarrow Q_\ssd, \qquad Q_\ssa \leftrightarrow 2 N,
\eeq
with $Q_\sse$ and $Q_\ssm$ fixed. That is, duality interchanges the
mass with the dilaton charge, as well as interchanging
the axion charge with the Taub-NUT parameter.

\ref\chchange{A. Giveon and M. Ro\v cek, \npb{380}{92}{128}.}

We stress that since this result relies just on the asymptotic form for
the fields involved, it is completely general and applies equally well to
exact string solutions as to lowest-order ones. One surprise is that the
relations we have derived amongst the various charges depend
on the asymptotic value of the electrostatic potential, $A_t$
--- we discuss this in more detail below. Our present results assume
that $A_t$ vanishes like $1 / r$ for large $r$.
As we shall see later,
if this large-$r$ behaviour is relaxed --- with $A_t$ approaching a
nonzero constant for large $r$ --- then, for example, $Q_\sse$ need not be
invariant under duality. (Duality transformations for which $Q_\sse$ is
not invariant have been considered for two-dimensional configurations
in Refs.~\chchange\ and \duality.)

\ref\gibbons{G.W. Gibbons and K. Maeda, \npb{298}{88}{741};\bk
D. Garfinkle, G. Horowitz and A. Strominger, \prd{43}{91}{3140}.}

A nice illustration of these general relations that has been considered in
the literature is the case of the magnetically charged dilatonic black
hole in four dimensions \gibbons, for which the dilaton charge, $Q_\ssd$, turns
out to be related to the mass, $M$, and the magnetic charge, $Q_\ssm$, by
$Q_\ssd=Q_\ssm^2/ (\GN M)$.  Here (assuming a gauge for which $A_t = 0$)
the dual solution is obtained from the original one by everywhere making
the replacement  \fandx:
\label\dmbhduality
\eq
2\GN M \to 2 \GN \twi{M} = {Q_\ssm^2 \over 2 \GN M} , \qquad \hbox{and}
\qquad {Q^2_\ssm \over 2 \GN M} \to {\twi{Q}^2_\ssm \over 2 \GN \twi{M}}
= 2 \GN M.
\eeq
Notice that eqs.~\dmbhduality\ --- together with the expression for $Q_\ssd$ in
terms of $Q_\ssm$ and $\GN M$ --- is a special case of the general
transformation rules derived here: $(2 \GN M, Q_\ssd, Q_\ssm) \to (2 \GN
\twi{M}, \twi{Q}_\ssd, \twi{Q}_\ssm) = (Q_\ssd, 2 \GN M, Q_\ssm)$.
In this particular example, the naked singularity domain, $Q^2_\ssm
> 2 \GN^2 M^2$, is mapped to the black hole domain, $Q_\ssm^2 < 2
\GN^2 M^2$, similar to what happens in two-dimensional examples.
The extreme case $Q_\ssm^2=2 \GN^2 M^2$ is a selfdual solution.

\section{The Dilaton--Metric Case in Detail}

It is instructive to work out the previous general manipulations in
a concrete example. We do so here using the explicit solution for the
dilaton--metric system to lowest-order in $\ap$. The results we
obtain lead us to consider in some detail the dualization
of the Schwarzschild black hole, including its nonleading $\ap$ corrections.

\subsection{Duality and the Lowest-Order Four-Dimensional Solutions}

We record here, for future use, the dilaton--metric string
solutions to the lowest-order in $\ap$ string equations in four
dimensions. (The higher-dimensional case is treated in the appendix.)
The relevant field equations arise from varying the
low-energy action of eq.~\leaction, and are given by
\label\eqofmotion
\eq \eqalign{
R_{\mu\nu}(g) &= {1\over 2} \; \nabla_\mu \phi \nabla_\nu \phi \cr
\nabla^2 \phi &=0 .\cr}
\eeq

\ref\fifties{H. Buchdahl, \pr{115}{59}{1325};\bk
A.I. Janis, D.C. Robinson and J. Winicour, \pr{186}{69}{1729}.}

The static, spherically-symmetric and asymptotically-flat solutions to these
equations are \fifties, \bmq\ given by the ansatz of eq.~\sssemetric\ with:
\label\answer
\eq \eqalign{
f = {1 \over g} &= \left( 1 - {\ell\over r} \right)^\delta \cr
h &= r^2 \left( 1 - {\ell\over r} \right)^{1-\delta} \cr
e^\phi &= \left( 1 - {\ell\over r} \right)^\gamma \cr}
\eeq
where $\ell$, $\delta$ and $\gamma$ are arbitrary constants,
apart from the one condition $\delta^2 + \gamma^2 = 1$.
The mass and dilaton charge of these solutions are given by $2 \GN M
= \delta \ell $ and $Q_\ssd = \gamma \ell$. For $\ell > 0$ the choice
$(\delta,\gamma)=(1,0)$ corresponds to the positive-mass Schwarzschild
solution, and $(\delta,\gamma)=(-1,0)$ also yields the Schwarzschild solution
up to a coordinate transformation, but with a negative mass.

Notice that when $\gamma \ne 0$, and so $Q_\ssd \ne 0$,
these field configurations contain
a curvature singularity at $r=\ell$, as can be
seen from the equations  of motion:
\label\singular
\eq \eqalign{
R(g) &= \hf \; (\nabla \phi)^2 \cr
&= {\gamma^2 \ell^2 \over 2 r^4} \left( 1 - {\ell \over r}
\right)^{\delta-2} . \cr}
\eeq

We now ask how these solutions transform under duality. Applying
the rules of the previous section, the dual turns out to be given by
\label\dualanswer
\eq \eqalign{
\fbar = {1 \over \gbar} &=\left( 1-{\ell\over r} \right)^\gamma \cr
\hbarr &=r^2 \left(1-{\ell\over r} \right)^{1-\gamma} \cr
e^{\phibar} &= \left(1-{\ell\over r} \right)^\delta \cr}
\eeq
Thus the duality transformation simply interchanges $\delta$ and $\gamma$.
This is just what is required to ensure
that the mass and dilaton charge are exchanged according to the general result
$2 \GN M \leftrightarrow Q_\ssd$. The  special cases $\delta =
\gamma = \pm {1\over \sqrt{2}}$ are distinguished by being self dual.

The previous discussion shows that the only solutions for which $r = \ell$
is nonsingular are those for which the dilaton charge, $Q_\ssd$, vanishes.
The positive-mass example of this is the solution for which $(\delta, \gamma)
= (1,0)$ --- \ie\  the positive-mass Schwarzschild black hole.
The dual solution, in this case,
has $(\tw{\delta} ,\tw{\gamma}) = (0,1)$, which is clearly one of the
configurations having a curvature singularity at $r=\ell$.

The result $\tw{\delta} = 0$ implies, in particular, that the mass, $\Mbar$,
of the dual solution vanishes, at least to lowest order in $\ap$.
In order to gain some
intuition for the physics of this dual configuration, we now compute
the dual mass to next order in $\ap$. We find the intriguing result that the
dual mass is given in terms of the original mass, $M$, by the relation
\label\dualmass
\eq
\Mbar = - \; {k \over \ap M},
\eeq
where $k$ is a known non-negative number. It is the purpose of the remainder of
this section to prove this result, and to determine $k$.
Before exploring the $O(\ap)$ corrections to
the lowest-order string solutions, however, it is first necessary to make a
short aside concerning the necessity of also modifying the
duality transformation rules at $O(\ap)$, which one might expect to
modify the physical charges at this order.

\subsection{$O(\ap)$ Corrections to the Duality Transformations}

The leading corrections to the effective action that are relevant for
modifying nontrivial gravitational solutions are \lestrings:
\label\fixaction
\eq
I=\int d^4x\,\sqrt{-G} e^\phi \left[ R(G) + (\nabla\phi)^2 +
{\lambda\over2} \; \ap R_{\mu\nu\rho\sigma}R^{\mu\nu\rho\sigma}
+\ldots\right]\ ,
\eeq
where $\lambda$ is, respectively, $\hf$ for bosonic, $\nth{4}$ for
heterotic, and $0$ for supersymmetric strings. There are also
higher-derivative terms that enter the effective action at this order
through the Lorentz Chern-Simons contributions to $H_{\mu\nu\rho}$, and
through field redefinition ambiguities. None of these terms are
relevant for the present discussion,  which is restricted
to static, uncharged, dilaton--metric configurations in four dimensions.

\ref\dualcorrections{
A.A. Tseytlin, {\sl Mod. Phys. Lett.} {\bf A6} (1991) 1721;\bk
C.P. Burgess, R.C. Myers and F. Quevedo, unpublished (1991).}

Although the duality transformation given by eqs.~\dualbusch\
is a symmetry of the leading order terms in the action, it is
{\it not} a symmetry of the $O(\ap)$ correction for general  curved
backgrounds.
The variation of the action turns out to be proportional to the field equations
themselves, however, indicating that the action can be made invariant by
modifying the duality transformation rules at $O(\ap)$
\dualcorrections.

For a sigma-model metric of the form,
\label\sssmetrictwo
\eq
dS^2=-F(r)\,dt^2+G(r)\,dr^2+r^2(d\theta^2+\sin^2\!\theta\,d\phi^2)
\eeq
the corrected duality transformation (associated with time translations)
which produces a symmetry of the action, including the $O(\ap)$
interactions is:
\label\dtrulestwo
\eq
\Fbar(r) = {1 \over F(r)}\left(1+ {\lambda\ap \over GF^2}\;
 \left({d F\over dr}\right)^2
\right), \qquad \Gbar(r) = G(r),
\eeq
and
\label\phiruletwo
\eq
\phibar(r) = \phi(r) + \ln \, |F(r)|- {\lambda \ap \over 2GF^2} \;
\left({d F\over dr}\right)^2\ \ .
\eeq
Of course, these transformations are understood to be applied
perturbatively in $\ap$, and one expects that
they receive further modifications at $O(\ap^2)$ and beyond.

In the present case where $F=1/G=1-{\ell \over r}$, we find
\label\newF
\eq
\Fbar = {1\over1-{\ell \over r}}\left[ 1+ \left( {\lambda \ap \over1-{\ell
\over r}}
\right) \; {\ell^2\over r^4} \right]
\eeq
and, since $\phi(r)=0$,
\label\newphi
\eq
\phibar=\ln\left(1-{\ell \over r}\right)- \left( {\lambda \ap \over1-{\ell
\over r}}
\right) \; {\ell^2\over 2r^4}\ \ .
\eeq

%
%
%
%

Thus we see that the $O(\ap)$ corrections to the duality
tranformation only introduce modifications in the asymptotic
metric and dilaton at $O({\ap \ell^2 \over r^4})$ in eqs.~\newF\ and
\newphi.\foot\coordchange{As was noted in Ref.~\dualcorrections, the
$\ss O(\ap)$ corrections to the duality transformations can be re-expressed
so as to preserve the previous duality transformation rule, $\ss \Fbar=1/F$,
provided that the transformation is combined with an $\ss O(\ap)$
coordinate change. This extra coordinate transformation also introduces
$\ss O({\ap\ell\over r^3})$ modifications into the metric.}
As a result, these modifications do not affect the leading asymptotic
behavior of the fields, including the gravitational and inertial masses,
and the dilaton charge. We expect this to be a general feature of any
higher-order corrections to the duality transformation. Any $O(\ap)$ terms
involve two derivatives of the fields, and so such terms cannot affect the
leading $O(1/r)$ asymptotic behavior. Hence the physical parameters,
such as the mass or dilaton charge, remain unchanged by these corrections.
This result should apply in higher dimensions as well.

\subsection{The $O(\ap)$ Corrected Solution}

The implications of the $O(\ap)$ corrections to the string equations
arising from eq.~\fixaction\ for the Schwarzschild black hole are given in
Refs.~\strbhone. The main new feature of these corrections is that
the dilaton can no  longer be held constant, since the (nonvanishing)
Riemann tensor provides a source for this field.
As a result, the coordinate singularity at the horizon of
the lowest-order solution generically becomes a
real singularity, unless the asymptotic form of the dilaton field is adjusted
to ensure that this does not happen. This tuning of the dilaton
corresponds to the freedom to introduce an arbitrary
dilaton charge into the solution, and with a particular choice for
$Q_\ssd$ at $O(\ap)$, the horizon singularity
is once again purely a problem with coordinates.
Refs.~\strbhone\ give the full form of the resulting solution
in any number of dimensions given this choice, but it suffices for our
own purposes to present here only the asymptotic form
for the result in four dimensions. This is  given by eqs.
\Gasymform, with constants given by:
\label\bhasymform
\eq
A = \ell + {23 \lambda \ap  \over 6 \ell} + O(\ap^2),
\qquad  B = \ell - {\lambda \ap  \over 6 \ell} + O(\ap^2), \qquad
Q_\ssd = - \; {2 \lambda \ap  \over \ell} + O(\ap^2);
\eeq
from which we also learn the form for the Einstein metric, namely
eq.~\gasymform, with:
\label\ebhasymform
\eq
2 \GN M = \Sca = \Scb = \ell + {11 \lambda \ap  \over 6 \ell} + O(\ap^2) .
\eeq
Recall that the constant $\lambda$ appearing here is $\hf$ in the bosonic
string, $\nth{4}$ in the heterotic string, and 0 in the superstring.

\ref\super{D.J. Gross and E. Witten, \npb{277}{86}{1};
M.T. Grisaru, A.E.M. van de Ven and D. Zanon, \plb{173}{86}{423};
\npb{277}{86}{388};\npb{277}{86}{409};
M.T. Grisaru and D. Zanon, \plb{177}{86}{347}.}
\ref\sustrbhap{R.C. Myers, \npb{289}{87}{701}.}

\subsection{The Duality Transformation}

The action of a duality transformation on the $O(\ap)$-corrected solution
of the previous section may be obtained as a special case of its action on
the asymptotic form for a general configuration. This is given in the present
instance by setting $Q_\ssa = Q_\sse = Q_\ssm = N = 0$  in eqs.~\smasymdual,
and using expressions \bhasymform\ and \ebhasymform\
for $Q_\ssd$ and $M$. This leads to the following expressions for the
asymptotic
form of the dual solution:
\label\smdasymform
\eq
\twi A = - \ell - {23 \lambda \ap  \over 6 \ell} + O(\ap^2),
\qquad \twi B = \ell - {\lambda \ap  \over 6 \ell} + O(\ap^2),
\qquad \twi Q_\ssd = \ell + {11 \lambda \ap  \over 6 \ell} + O(\ap^2);
\eeq
and so
\label\edasymform
\eq
2 \GN \twi M = \twi \Sca = \twi \Scb = - \; {2 \lambda \ap  \over \ell}
+ O(\ap^2).
\eeq

A comparison of eqs. \ebhasymform\ and \edasymform\ gives the
advertised relation between the masses of the dual solutions:
\label\mdual
\eq
\Mbar = - \, {k \over \ap M} + O\!\left({1\over\ap^2M^3}\right),
\eeq
where $k$ is a dimensionless constant, given explicitly  by:
$k = \lambda \ap^2 / 2 \GN^2 = 2\lambda e^{2\phi_0}$. This
result is similar to what has been found in two-dimensional examples
\moretwodbh. The mapping of positive to negative masses under
duality was also found for other circumstances in Ref.~\welch.
Notice, though, that this mapping is {\it not} a generic property,
since it fails for those solutions in eq.~\answer\
for which $\delta$ and $\gamma$ have the same sign.

For the superstring, where $\lambda=0$, even the $O(\ap)$ corrections are
insufficient to determine the sign of the mass of the dual configuration.
For this particular case, the leading corrections to the low energy
string equations are
$O(\ap^3)$ \super, and one must work to this order to determine the
mass of the dual solution. The implications of these corrections for
black hole solutions have also been determined \sustrbhap.
With these results in hand and repeating the previous analysis,
one finds that
\label\sstringcase
\eq
\Mbar = - 3 \zeta(3) \; { \ap^3 \over (2 \GN)^6 M^5} =
 - 3 \zeta(3) \; { e^{6 \phi_0} \over \ap^3 M^5},
\eeq
where $\zeta(z)$ denotes the Riemann zeta function.
Once again the dual mass is negative.

For the two-dimensional black hole it is also true that the positive-mass
black hole is dualized to a negative-mass solution. In the two-dimensional
case, however, this relation between positive- and negative-mass solutions
is related to the mapping under duality of the asymptotically flat region
to the region `beyond' the singularity at $r=0$ ---
both of which are incorporated as different sectors of
a single conformal field theory. A similar statement for the
higher-dimensional case of interest here would be very interesting but,
unfortunately,
must await more detailed knowledge concerning the solutions, and the
corresponding conformal field theories.

\section{The (In)Equivalence of Dual Solutions}

An important question for these investigations
is whether or not dual solutions describe
identical physics. This question is  crucial if duality transformations are to
be used
to infer the properties of string solutions which would otherwise not be
amenable to direct analysis. One instance for which there is good evidence
for this equivalence is when the orbits of the symmetry on which duality is
based are compact \compacteq. In this case the two dual solutions have
been argued to represent exactly the same conformal field theory, and so
to represent identical string physics.

Unfortunately, this result cannot be directly applied to the present case,
since the time-translation symmetry, $t \to t+c$, on which our dualization
is based, is not compact. In fact, our result that positive-mass solutions get
taken to negative-mass ones might make one wonder whether such
`noncompact' duality transformations need relate physically equivalent
situations at all. The present section is devoted to further exploring this
issue
within the concrete setting of the dilaton--metric configurations that are
discussed above.

We do this exploration in two steps. First, we compute the scattering of
massless particles by the four-dimensional dilaton-metric configurations.
Part of our purpose for doing so is to
verify the physical significance of the result $M_\ssg \to \Mbar_\ssg =
- k/(\ap M_\ssg)$, for the gravitational mass, $M_\ssg$, by rederiving it
within the context of a directly observable process.  After all, one might
wonder whether the existence of the long-range dilaton field might
alter particle propagation in such a way as to compensate for the change
in the gravitational mass that occurs in passing from a particular
dilaton--metric
background to its dual. We find that this is not how things work, since the
dilaton drops out of the propagation equations for massless particles in the
geometric optics approximation.
Our second step is to examine several arguments in favour of the
inequivalence of such solutions.

\subsection{Massless Particle Scattering}

Consider, then, using a massless string state as a test particle with which
to probe the background metric and dilaton fields. We wish to compute
the scattering of such a state by these field configurations, with
the goal of comparing the result for backgrounds that are related by
duality.

The first question we must answer concerns how the background fields
affect the motion of such test particles. The classical propagation of a
closed string through a given background is
determined by finding
the string world-sheet which is a stationary point of the
Polyakov string action, and which has the topology of a cylinder. This
world sheet action depends on the background dilaton field configuration,
$\phi(x)$, through the coupling term \lestrings:
\label\dilatonterm
\eq
- \; {1 \over 8 \pi} \int d^2\sigma \sqrt{-h} \; \phi[x(\sigma)] \; \Scr,
\eeq
where $x^\mu(\sigma)$ are the coordinates of the string world sheet,
$h = \det \, h_{ab}$ is the determinant of the world-sheet metric,
and $\Scr$ is its intrinsic curvature. But the cylindrical worldsheet
of a classically  propagating string is conformally flat, and so $\Scr$
can be taken to everywhere vanish. This implies that the dilaton is
completely irrelevant for describing such classical string motion.

Furthermore, for a test string for which only massless modes are excited,
it is a good approximation to neglect the
dependence of $x^\mu(\sigma)$ on the spatial coordinate along
the string, and to consider it to be simply a function of the proper
time, $\tau = \sigma^0$, of its centre of mass. In this case, the classical
equations which determine $x^\mu(\tau)$ reduce to the equations
for a null geodesic of the background spacetime metric. Since null geodesics
are not affected by conformal rescalings of the metric, it
is immaterial whether this null geodesic is computed using the
sigma-model metric, $G_{\mu\nu}$, or the Einstein metric, $g_{\mu\nu}$.

\ref\mende{P.F. Mende, lecture at the Workshop on String Quantum
Gravity, Erice, 1992, preprint Brown-HET-875, hep-th/9210001;
R. Lafrance and R.C. Myers, ``Gravity's Rainbow,'' preprint
McGill-94-45, hep-th/9411018.}

An alternative argument to the same end goes as follows. One could instead
think of test-string propagation in terms of the four-dimensional effective
theory of eq.~\leaction, which describes the low-energy propagation of
massless string states. In this case, the classical propagation of, say,
a photon is described, to lowest order in $\ap$, by the Maxwell equations
$\nabla^\mu(e^\phi \; F_{\mu\nu}) -{1\over2}\epsilon_{\nu\mu\rho\sigma}
(\nabla^\mu a)F^{\rho\sigma}= 0$. For photons with wavelengths
that are much longer than the string scale, but also much smaller
than the radius of the background curvature or the scale of variation
of the background dilaton and axion fields, these equations may be solved
in the geometrical-optics approximation \weinberg\mende. This
leads, once again, to the null geodesics which are the same for
the sigma-model and Einstein metrics.

Using either line of argument, it is clear that the background dilaton
field cannot affect the scattering of massless test strings in the regime
we are considering. Consider, therefore the scattering of such a massless
string state using the sigma-model metric of eq.~\sssmetric:
\label\unlabelled
\eq
dS^2 \equiv G_{\mu\nu} \, dx^\mu dx^\nu = - F(r) \, dt^2 + G(r) \, dr^2 + r^2
(d\theta^2 + \sin^2\theta \, d\varphi^2) ,
\eeq
where we have performed a coordinate transformation to ensure $H(r) = r^2$.
It is a straightforward calculation \weinberg\ to find the net scattering
angle, $\Delta \varphi$, through which an incident null geodesic
deviates while passing through such a gravitation field. In terms of the radial
coordinate, $r_0$, of the point of closest approach of the null geodesic
to $r=0$, the scattering angle is given by:
\label\scatangle
\eq
\Delta \varphi + \pi = 2 \left| \int_{r_0}^\infty G^{1/2}(r) \left[ \left(
{r \over r_0} \right)^2 \left( {F(r_0) \over F(r) } \right) - 1 \right]^{-1/2}
{dr \over r}\, \right|.
\eeq

If $r_0$ is chosen large enough to justify using the asymptotic large-$r$
expansions: $F(r) \simeq 1 - A/r$ and $G(r) \simeq 1 + B/r$,
then eq.~\scatangle\
reduces to
\label\simpang
\eq
\Delta \varphi \simeq {2 \xi \over r_0} ,
\eeq
where $\xi = \hf \, (A+B)$. For the nonsingular, Schwarzschild-like
solutions $\xi$
is given explicitly by $\xi \simeq \ell + 11 \lambda \ap  /(6 \ell) +
O(\ap^2)$.

After a duality transformation we have $\Fbar = 1/F$ and $\Gbar = G$. So the
scattering of massless string states in the dual metric is again given, in the
large-$r_0$ limit,  by eq.~\simpang\ but with the replacement
\label\dualrepl
\eq
\xi \to \tw{\xi} \equiv \hf \; (\Abar + \Bbar) = \hf \; (B - A).
\eeq
For the metric that is dual to the nonsingular, Schwarzschild-like, one, we
therefore have $\tw{\xi} \simeq - 2\lambda \ap /\ell + O(\ap^2)$. Notice
that $\xi$ and $\tw{\xi}$ therefore have opposite signs, and so predict the
deflection of an incident massless string state into the opposite directions.
Notice also that the relation between $\xi$ with $\tw{\xi}$ is precisely the
same as that obtained earlier between $\Sca$ and $\abar$, or $\Scb$
and $\bbar$, \ie: $\ell \to - 2\lambda \ap  / \ell$, reproducing the result
$M \to -k/(\ap M)$. For the self dual metric
$(\delta,\gamma)=({1\over\sqrt{2}},{1\over\sqrt{2}})$ we can see
that $A=0$ and so $\xi = \tw{\xi}$ as it should, but for all the
other values of $\delta$ and $\gamma$ we find  $A\neq 0$ and
therefore $\xi \neq \tw{\xi}$. So in general it appears that the dual
backgrounds lead to distinct physical results. 

\subsection{Equivalent or Not?}

Finally then, we attempt to address the question
of whether or not  string solutions that are related by a
duality transformation based on a noncompact symmetry
are physically equivalent. Superficially, the calculation
of massless-string scattering just described seems to indicate
that dual solutions represent physically inequivalent
configurations, although as we discuss below, this result is not
in itself completely 
conclusive. We nevertheless believe configurations that are
related by dualities based on time translation could be inequivalent,
at least in four or more dimensions. We list, in this section,
the arguments for,  and against, the equivalence of the dual
configurations we have considered.

\topic{1. Two Dimensional Black Holes}
If the dual of the black hole solution really is physically different
than the black hole solution itself, then this conclusion
runs counter to the conventional wisdom about two-dimensional
black holes in string theory.\foot\nocont{There is not a flat contradiction,
since the above scattering analysis cannot be applied to the
two-dimensional examples. See item {\it 4} below, however.}
The two-dimensional solutions arise from a
gauged $SL(2,\IR)/U(1)$ WZW model \twodbh. There are
arguments that in this case target-space duality relates equivalent
conformal field theories \kiritsis, despite
the fact that the masses of the two solutions have
opposite signs \moretwodbh.

The relation between the dual configurations is quite subtle, in
the WZW model, since the total space is considered to incorporate several
copies of (the maximal analytic extension of) the black hole and the
negative-mass naked singularity \moretwodbh, \paulandme. Duality
then relates different asymptotic regimes of the extended solution.
Unfortunately, a similar understanding in the present case
is not yet possible, since the conformal field theory representation of
the four-dimensional solutions is unknown. As a result there is no
guidance as to how negative- and positive-mass spaces might be subsumed
within some larger string solution. On a related point, we note that
the higher dimensional 
solutions with a nonvanishing dilaton charge become complex in
the region $0\le r<\ell$, and so are more pathological in this region than
is the usual black hole for $r<0$.  Hence, for a fixed dilaton charge, the
positive and negative mass solutions are not naturally linked in a single
(real) geometry.

\topic{2. Momentum and Winding Modes}
There is another argument that can be made against using
the calculation of the scattering of massless string states to address
the equivalence of dual configurations. This argument is based
on the fact that experience with the action of duality on a torus shows
that the momentum and winding modes in the symmetry direction should be
exchanged by duality. Thus, any physical equivalence between the various
black hole solutions would relate the physics of momentum carrying states
to that of the winding states in the dual solution.

\ref\bars{I. Bars and J. Schulze, preprint USC-94/HEP-B1,
hep-th/9405156 (1994).}

In all of our four-dimensional solutions, however, the symmetry
direction is timelike, as well as being noncompact. Both of these
properties make it difficult to introduce the dual
`winding' states in a sensible way.
Thus, in our comparison of dual solutions, we considered
the scattering of momentum-carrying states for both spaces. Within the
limitations of our present understanding, this seems to be the only physically
relevant comparison that can be made.\foot\zeromass{Further we note
that following the example of the torus, one expects the massless states to
remain massless under duality}
Perhaps an analysis along the lines of
Ref.~\bars\ would provide some new insight.

The implications of duality for CFT's in the case of noncompact symmetries
has been studied in some detail in Ref.~\aabl. These authors found that the
images under duality of ordinary momentum-carrying modes are a class
of {\it nonlocal} vertex operators (`vortex lines').
Thus, although
this construction can be taken to define an equivalent, dual, conformal field
theory in the noncompact case, the resulting theory in general differs from the
usual string quantization, as defined as the sum over local string embeddings
in the corresponding spacetime.  This distinction is probably made clearest
by considering the example of standard string propagation in flat Minkowski
space. In this case the image spacetime under duality --- based on one of the
noncompact symmetries, such as translations --- is also flat Minkowski space.
But since the dual string theory constructed using such a duality
transformation
includes the nonlocal `vortex line' vertex operators, it is {\it different}
from the starting theory in which no such operators appear.

\ref\handh{J.B. Hartle and S.W. Hawking, \prd{13}{76}{2188}.}

\topic{3. Euclidean Solutions}
One might hope to better address these issues by dualizing the
euclidean version of our family of dilaton-metric solutions
in eq.~\answer. (One could also consider the general solutions
presented in Ref.~\bmq.) In this case, for the (nonsingular) black
hole geometry, there is a natural periodic identification of the euclidean
time direction --- $t \simeq t + \beta$ with $\beta = 2 \pi \ell$ --- which
avoids a conical singularity  at $r=\ell$, and so produces a
completely smooth manifold \handh. Of course, such an identification
cannot cure the curvature singularities in the solutions having
nonzero dilaton charge,  and so in this case there is no natural period
which suggests itself for the euclidean time direction. It seems likely
that the spectrum of the string in the background of a euclidean black
hole will be identical to that obtained
with the dual solution, if, in the latter, euclidean time is also identified
with the inverse period --- \ie, $\beta \to \tw\beta \sim \ap / \beta$.
This expectation arises because in this case the symmetry
direction is compact. Further one expects that the same result holds
for dual euclideanized solutions with arbitrary dilaton charge.

This construction does not imply a similar equivalence for the Minkowski
signature vacua. The Minkowski-signature problem
which corresponds to the euclidean system with time-period $\beta$ is a
system in a heat bath of temperature $T=1/\beta$. Thus, the duality would
relate
two Minkowski-signature systems whose temperatures are related by $\twi{T}
\sim 1/(\ap T)$. Implicit in our previous discussion is that we are
considering the physical equivalence of the zero-temperature vacua for
the dual solutions. 
The reciprocal relationship between the dual temperatures though
implies that such a comparison can not be made here, since as
$T \to 0$, the temperature diverges for the equivalent
dual heat bath. 

Another attempt might be to make a periodic
identification of time directly in the original Minkowski-signature
solutions. For the static solutions, there is no natural criterion for
selecting the time period,  but for a solution with a nonvanishing
NUT parameter as in eq.~\TNUTansatz, a natural period for the time
coordinate, $8\pi N$, is suggested by the requirement that
the $\theta=0$ and $\theta=\pi$ axes be free of conical singularities
\misner.\foot\footnotefest{If this choice is made, then since
$\ss Q_\ssa \leftrightarrow 2 N$
under duality, the periodicity of two dual solutions would in general
be completely unrelated. They would not, in particular, be related
in the usual `dual' way \ie\ $\ss \tw\beta \sim \ap / \beta$.}
The resulting solutions would in any case display rather
pathological characteristics, similar to those of the
Taub-NUT or Misner spaces \misner.
The interpretation of  winding modes in time-like directions would also
remain unclear, and so we did not pursue this approach here.

\topic{4. Background Gauge Fields}
There is another argument which we bring forward against the idea
that the duality tranformations in eqs.~\dualbusch and \dualgaugx\
always relate physically equivalent solutions.
This argument has the advantage that it addresses the potential
objections of arguments {\it 1} and {\it 2} above.

Notice that if we consider a general solution
which includes background gauge fields,
then the duality transformations become more complicated
(see eq.~\dualgaugx\ and the surrounding discussion). As was
mentioned earlier, our results concerning the action
of duality on the mass and the various charges rely crucially on the assumed
vanishing of ${A}_t$ like $1 / r$ at infinity.
This assumption is crucial because it is required to ensure that the
contributions of $A_\mu$ to the dual fields --- which
are typically proportional to $A_t^2$ --- fall off like $1/r^2$, and so
do not affect their leading asymptotic behavior.

Before proceeding to our final argument, let us first consider in detail
the case where $A_t$ approaches a constant asymptotically,
\label\newguage
\eq
A_t=2v+{Q_\sse\over r}+\cdots,
\eeq
which leaves the asymptotic behavior of the electric field \emaform\
unchanged. In order to keep the asymptotic form for the field strength,
$H_{t\theta\varphi} = - \; Q_\ssa \sin\theta + \cdots$, unchanged the
asymptotic behavior of the Kalb-Ramond field must also be modified,
because of the Chern-Simons contribution to $H_{\mu\nu\rho}$ (\ie\
$H=dB-{1\over4}A\,dA$). The required asymptotic form is:
\label\newbmunu
\eq
B_{\varphi t} = (Q_\ssa+{v\over2} \, Q_\ssm)\, \cos\theta + \cdots\ .
\eeq
After performing the duality transformation, one finds that
asymptotically  $\Gbar_{tt}\rightarrow-1/(1-v^2)^2$.
Rescaling the time coordinate, $t= (1-v^2) \, \tt$, recovers
the desired asymptotic form, $\Gbar_{\tt\tt} \, \rightarrow -1$.
Notice that with this choice, the asymptotic value of the gauge
potential is preserved, \ie, $\Abar_{\tt}\rightarrow2v$.
It is also necessary to shift  the dilaton by a constant in order to
recover $e^\phi\rightarrow1$ as $r\rightarrow\infty$.
Now it is straightforward to calculate the new physical parameters
which characterize the dual solution, and one finds
\label\longeq
\eq
\eqalign{
2{\twi{\GN}}\Mbar&=
{1\over1-v^2}\left[Q_\ssd-2v^2\GN M-v Q_\sse\right]\cr
\Qbar_\ssd&={1\over1-v^2}\left[2\GN M-v^2Q_\ssd+v Q_\sse\right]\cr
2\Nbar&={1\over1-v^2}\left[Q_\ssa- 2v^2N\right]\cr
\Qbar_\ssa&={1\over1-v^2}\left[(1-2v^2)2N+v^2Q_\ssa\right]\cr
\Qbar_\sse&={1\over1-v^2}\left[(1+v^2)Q_\sse+2v(2\GN M-Q_\ssd)\right]\cr
\Qbar_\ssm&=Q_\ssm+{v\over1-v^2}\left[2N-Q_\ssa\right]\ \ .\cr}
\eeq
In deriving these results, we have assumed that $v^2<1$.
Notice that these expressions are singular as $v\rightarrow\pm1$,
and further that the limit $v\rightarrow0$
recovers the previous results of eq.~\dualsum.
Further, in the first line above, we write $\twi{\GN}$ to emphasize
that Newton's constant is modified since we made a constant shift
of the dilation field so that after the duality transformation
it still vanishes asymptotically.

We remark {\it en passant} that some of the consequences of
eqs.~\longeq\ can be cast in a form which is similar to the $v=0$ case:
\label\simpeq
\eq
\eqalign{
2\GN M+{v\over4} \, Q_\sse&\leftrightarrow Q_\ssd-{v\over4} \, Q_\sse\cr
2N+{v\over2} \, Q_\ssm&\leftrightarrow Q_\ssa+{v\over2}\,Q_\ssm\ \ .\cr}
\eeq
Another simple consequence is
\label\simpeqq
\eq
(2{\twi{\GN}}\Mbar-\Qbar_\ssd)^2-\Qbar_\sse^2=
(2\GN M-Q_\ssd)^2-Q_\sse^2\ \ .
\eeq

Now comes the main point.
The result that the physical charges depend on the asymptotic
value of $A_t$, may bear on the physical equivalence of
backgounds related by the duality transformation we have
considered. The starting observation is that
an arbitrary constant may be added to $A_t$ via a gauge
transformation. One could therefore imagine producing a family of
gauge-equivalent --- and so physically equivalent --- solutions to the
string equations, all differing only by their asymptotic values for $A_t$
and $B_{\varphi t}$. Under duality, this family of gauge-equivalent
solutions yields a family of dual solutions which all
have different physical characteristics, as shown in eq.~\longeq,
 since the asymptotic forms for these fields depend on $2v$,
the asymptotic value of $A_t$. Since this dual
family of solutions have different charges, they cannot be
physically equivalent.\foot\thisiswhy{This could be confirmed, for instance, by
studying massless-particle propagation in the geometrical optics limit,
as in section 5.1.  Since this propagation is again given by the null geodesics
of the metric (as may be seen by examining the field equations of the
low-energy effective theory), it is completely controlled at large
distances by the solution's mass, $\ss M$. }\ But this inequivalence
is inconsistent with the gauge equivalence of the original family
of backgrounds, together with any hypothetical equivalence
between each solution and its dual. It would appear that, at
best, the original family of gauge-equivalent backgrounds
could only be physically equivalent to one of the dual solutions.

Notice that in the case of a compact symmetry, the same argument
cannot be made, since in this case no gauge transformation would be
possible to shift $A_0$ by an arbitrary constant value. (The subscript `0'
here is meant to generically indicate the symmetry direction, rather than
time.) Instead, backgrounds with a different constant
values for $A_0$ would be considered as
physically distinct, since they are characterized by distinct
Wilson lines.\foot\discretecase{A discrete version of the preceding
argument might be possible even in this case, based on the large gauge
transformations which can shift the Wilson lines by a discrete amount.
It would be interesting to investigate the physical equivalence of such
theories in more detail.}\ Further, let us note that
in the noncompact case, the gauge parameter implementing
the shift of $A_t$ is divergent as $t\rightarrow\pm\infty$,
and does not vanish asymptotically as $r\rightarrow\infty$.
Despite these apparently problematic features, gauge
potentials $A_t$ related by a constant shift are usually considered
physically equivalent.

This argument implies that one of the following three conclusions must
be true. Either: ($i$) Dual field configurations need not be physically
equivalent; ($ii$) Constant shifts of the electrostatic potential, $A_t$,
change the physical content of the theory; or ($iii$) The
transformations described by eqs.~\dualbusch\ and \dualgaugx\ do not
correctly describe
duality transformations for arbitrary asymptotic values of $A_t$.
\endtopic

\section{Conclusions}

The purpose of this paper has been to investigate how duality acts on
string black-hole solutions in four (and higher) dimensions. We have
obtained the following results:

\topic{(1)}
We have investigated the action of duality (based on the symmetry of
time-translation) on asymptotically flat field configurations.
By considering the action of duality on the asymptotic form of a general field
configuration in four dimensions, we have been able to show that it simply
interchanges the configuration's mass and dilaton charge, as well
as interchanging the axion charge with the Taub-NUT parameter.
The electric and magnetic charges remain invariant. We emphasize that
this result is completely general, and does {\it not} hold only for solutions
of the leading order string equations. This generality of our result follows
since we have based our analysis purely on the asymptotic behaviour of
the fields at $r \to \infty$, and this asymptotic behaviour
should be unaffected by higher-derivative string corrections to the
equations of motion.
We illustrate this general conclusion in the text, by explicitly
constructing the action of duality for a general class of dilaton-metric
solutions in four dimensions. In the case of
an electrically charged black hole, our result that
$Q_\ssd$ and $M$ interchange under duality reproduces some older results
of ref. \fandx. (The case of higher-dimensions is treated in an Appendix,
where similar results are derived.)

It should be borne in mind that our conclusions
regarding the exchange of various physical parameters
under duality rely on the very plausible assumption
that any $O(\ap)$ corrections to the transformation
rules do not affect the values of these charges.
Certainly this was found to be the case for the pure dilaton-metric
solutions in section 4. The general argument presented there
was that dimensional considerations indicate that
the $O(\ap)$ corrections must involve two derivatives of
fields, and so they could only modify the asymptotic form of the
dual solutions at $O(1/r^3)$. This would not affect the
physical parameters of interest here.
(A similar result should apply for the analysis in higher
dimensions given in the Appendix.)

Of course, the preceding discussion does make the assumption that
the gauge potential $A_t$ vanishes asymptotically. If, instead, the
potential approaches a constant, \ie, $A_t\rightarrow 2v$, the action of
the duality transformations on solutions is still easily described
in terms of the transformation of the physical charges. The results
in this case, eq.~\longeq, are somewhat more involved though.
One may ask whether our results also depend on the particular
asymptotic values which we have chosen for the other fields. In fact,
they don't. For example, one could have the asymptotic
value $G_{tt} \rightarrow -Z^2$. A simple scaling of the time
coordinate would reduce this value to $-1$, as is used in section 2.2.
In principle, we have two options to consider: we can apply the duality
transformation after scaling time (yielding the results
discussed above), or before. In the latter case, one still finds
that a combination of scaling time and shifting the dilaton by
a constant yields the same dual background that results
from the first choice. Thus this asymptotic value of $G_{tt}$ has
no effects on the physics of the dual background. In other fields,
the asymptotic values are equally inconsequential for the action of
the  duality on the physical charges, although the precise implementation
may be slightly more involved. For example: via a (time-independent)
diffeomorphism, a metric component $G_{tr}$ could be induced. After the
duality transformation \dualbusch, one would find a new $B_{tr}$ component
in the Kalb-Ramond field, but the latter could be removed via a
gauge transformation of this field, \ie, $B_{\mu\nu}\rightarrow
B_{\mu\nu}+\partial_{[\mu}\omega_{\nu]}$.\foot\ilovethem{This
is what happens in the case
of duality for the Schwarzschild black hole expressed in terms
of the Eddington--Finkelstein metric. As was noticed in
Ref.~\fandx, the dual metric obtained in this case is identical to
what is found using the Schwarzschild metric, but it is supplemented
by a nonvanishing $\ss B_{\mu\nu}$. The equivalence of the dualization
in Eddington-Finkelstein coordinates is then seen by noticing that
the resulting $\ss B_{\mu\nu}$ can be gauged to zero
by a transformation $\ss B_{\mu\nu}\rightarrow
B_{\mu\nu}+\partial_{[\mu}\omega_{\nu]}$.}

The asymptotic value of $A_t$ seems to be distinct in that originally
it can be removed via a gauge transformation, but after the duality
transformation the presence of this parameter in the other fields is no
longer pure gauge, as can be seen by its effect on the physical charges in
eq.~\longeq.

\ref\nats{M. Natsuume, \prd{50}{94}{3949}.}

\topic{(2)} The special case of the four-dimensional `Schwarzschild-like'
solutions --- \ie\ those that are nonsingular at the Schwarzschild radius ---
is particularly interesting. In this case a  black hole of mass
$M$ is mapped onto a singular configuration whose mass vanishes to the
lowest order in $\ap$. By evaluating the next-to-leading corrections in $\ap$
to this solution, we are able to obtain the mass, $\Mbar$, of this dual
configuration. We find the intriguing result $\Mbar = - k/(\ap M)$, where
$k = \lambda \ap^2 / 2 \GN^2 = 2\lambda e^{2\phi_0}$. $\lambda$ here
is $\hf$ in the bosonic, and $\nth{4}$ in the heterotic string.

For the superstring, it happens that $\lambda = k = 0$ in the above formula,
and so the results of the $O(\ap^3)$ corrections to the black-hole
configuration are required in order to determine the dual mass.
We find, in this case, the negative mass: $\Mbar = -  \, k' / ( \ap^3 M^5)$,
where  $k' = 3 \zeta(3) \, e^{6 \phi_0}$.

With these results and the analysis given in the Appendix,
we have found then that, for all $d\ge4$, the nonsingular (uncharged)
black hole solution gets mapped onto a solution having negative mass,
just as for two dimensions \moretwodbh.
This is {\it not} a general feature of the dual configurations, as
may be seen from the four-dimensional solutions for which
both the mass and dilaton charge can be positive.

We would like to remark here that higher derivative terms in the
effective string equations do induce $\ap$ corrections to
the Schwarzschild metric and hence
the charges as found in equations \bhasymform\ and \ebhasymform.
These corrections arise, as discussed in Ref~ \strbhone, from
the condition of maintaining a regular horizon
in the solution to the $\ap$ corrected field equations.
An analogous result was obtained for the charged
dilatonic black hole in Ref.~\nats, where the condition of maintaining
a regular extremal horizon is imposed. This $\ap$ dependence of the
charges should not
be confused as being in contradiction with our statement that duality
makes no $\ap$ corrections for the corresponding charges. Given a
solution for which the physical charges have a complicated $\ap$
expansion, the action of duality is still to simply trade the
mass with the dilaton charge, as well as the axion charge with the
Taub-NUT parameter, as in eq.~\dualsum.

\topic{(3)}
We have also addressed the potential inequivalence of the two dual
solutions in the case where the symmetry is not compact. Using the
calculation of the scattering of massless string states we saw that
the dual solutions differed in their predictions. Starting from this
we presented several arguments which bear on the equivalence
of the dual configurations. In particular our final argument forces
us to choose one of the following three options: ($i$) dual solutions can be
physically inequivalent; ($ii$) constant shifts of the electrostatic
potential, $A_t$, can change the physical content of the theory; or
($iii$) although the transformation given in eqs.~\dualbusch\ and
\dualgaugx\ is a legitimate element of $O(d,d+p)$ \senn\ (and
so generates a new background solution), its identification as the
duality transformation rule with background gauge fields
is incorrect.
We further note that while our final argument of the last section
explicitly refers to four spacetime dimensions, it can easily be generalized
to other cases.

Setting aside the problems arising with background gauge fields,
it appears that Ref.~\aabl\ provides the most plausible
explanation for the physical equivalence of solutions related
by duality based on a noncompact symmetry --- namely, this
equivalence requires that one of the dual string theories
be quantized as vortex gas. In this unconventional string theory,
there are no local vertex operators carrying momentum in the symmetry
direction, rather one constructs non-local `vortex' operators carrying
`winding-number' for this direction. In comparing the dual
theories, one must keep in mind this distinction --- \eg, in the
two-dimensional black hole, the propagation and interactions
of tachyons near the
singularity are equivalent to the propagation and interactions
near the horizon of some dual vortex states, rather than of
tachyons. This line of argument
suggests that for a given configuration of background fields,
there may be more than one consistent string theory. It would
be interesting to determine if there are consistent quantization
schemes which incorporate both the local and non-local states
for noncompact directions,
or if this ambiguity only arises in coincidence with
background symmetries. In the present context, one unsettling
aspect of this approach is that the symmetry direction is the
time direction, and so the interpretation of the vortex theory
is somewhat perplexing. Certainly much more work is needed in order
to clarify these issues.
\endtopic

\ref\AAL{E. Alvarez, L. Alvarez-Gaum\'e and Y. Lozano, ``On
Non-Abelian Duality'', preprint CERN-TH-7204/94, hep-th/9403155.}

There are a number of other directions in which the work presented here
could be pursued. One would be to examine the transformations
given in eqs.~\dualbusch\  and \dualgaugx\ as
duality transformation rules with background gauge fields
for a compact symmetry. Of particular interest are the effects of large
gauge transformations.
Another interesting extension of our results may be to examine
the implications of duality based on the rotational symmetries of the
solutions we have considered. This could be done using either abelian
duality, based on constant shifts of the angular coordinate, $\varphi$,
or non-abelian duality, possibly following  Ref.~\AAL, using the entire
rotational group.

\bigskip
\centerline{\bf Acknowledgments}
\bigskip

C.B. would like to thank the Institut de Physique at the Universit\'e
de Neuch\^atel for its kind hospitality while this work was in progress.
We acknowledge useful conversations with E. Alvarez,
L. Alvarez-Gaum\'e, E. Kiritsis, Y. Lozano,
P. Page, J. Russo,  and A. Tseytlin.
This research was partially funded by the N.S.E.R.C.\ of Canada,
les Fonds F.C.A.R.\ du Qu\'ebec, and the Swiss National Foundation.

\bigskip

\vfill\eject
\appendix{A}{Duality in Higher-Dimensions}

We record in this appendix the action of duality on the lowest-order
static, spherically symmetric, asymptotically flat
solutions for spacetime dimensions $d > 4$. For generic
values of $d$, the nontrivial asymptotic fields only
include the metric, the dilaton and the gauge potential $A_t$,
and hence the solutions are characterized by their mass,
dilaton charge and electric charge.
The lowest-order solutions are given explicitly in Ref.~\bmq,
and we use these solutions (which are also valid for $d=4$)
as our starting point.

The metric in the Einstein frame may be written as
\label\newcharo
\eq
ds^2 = - {U^2\over W^{2(d-3)/(d-2)}} \; dt^2
+ W^{2/(d-2)}V^2 (dr^2 + r^2 d\Omega_{d-2})
\eeq
where $d\Omega_{d-2}$ is the standard line element on a unit $(d-2)$-sphere.
Recall that for arbitrary dimension $d$, the sigma model and Einstein
metrics are related by $G_{\mu\nu}=e^{-2\phi/(d-2)}g_{\mu\nu}$.
The solutions also include the following dilaton and gauge fields
\label\newchart
\eq
\eqalign{
e^{{\phi}}&=W/X\cr
A_t&=2v+\sqrt{x^2-1} \; \left( {1-U^2X^{2/(d-2)}\over W}
  \right) \ \ .\cr}
\eeq
These fields are written in terms of the following functions
\label\solutionn
\eq \eqalign{
W&={1\over2} \; \left(1+x+(1-x)U^2X^{2/(d-2)} \right)\cr
U^2 &= \left( {\beta\over\alpha} \right)^{2H} \cr
V^2 &= (\alpha\beta)^{2 /( d-3)} \left( {\alpha\over\beta}
\right)^{2H/(d-3)} \cr
X &= \left( {\alpha\over\beta} \right)^{K}, \cr}
\eeq
where
\label\moree
\eq
\alpha = 1+\left({\ell\over 4r}\right)^{d-3}
\qquad\qquad
\beta = 1-\left({\ell\over 4r}\right)^{d-3}.
\eeq
The constants $H$ and $K$ must satisfy $H^2+K^2 (d-3) / (d-2)^2 =1$.
These constants as well as $x$ parameterize the three physical
charges of these solutions: the mass, the dilaton charge
and the electric charge,\bmq
\label\gimasses
\eq
\eqalign{
2\GN M &= {A_{d-2}  \over2 \pi }
\left[H\left(d-2+(x-1)(d-3)\right)
-K(x-1){d-3\over(d-2)}\right]\left(\ell\over 4\right)^{d-3}\cr
Q_\ssd &= \left[2 K\left(1+{x-1\over d-2}\right)
-2H(x-1)\right] \left( {\ell \over 4} \right)^{d-3}\cr
Q_\sse &=\sqrt{x^2-1}\ (d-3)\left(4H-{4K\over d-2}\right)
\left({\ell\over 4} \right)^{d-3}\cr}
\eeq
where $A_{d-2}$ is the area of the unit $(d-2)$-sphere.
A fourth constant appears in the gauge potential which
is the asymptotic value of the gauge potential, \ie,
$A_t\rightarrow 2v$. For $K=0$, the solutions correspond
to charged black holes in which the surface $r=\ell/4$ is
a nonsingular event horizon.


Duality transformations map these solutions
amongst themselves. We begin by focusing our attention
on the dilaton-metric solutions with $A_t=0$, \ie,
$v=0$ and $x=1$. Then eq.~\gimasses\ shows that the
mass is proportional to $H$, and the dilaton charge, to $K$.
Under a duality transformation,
one finds $(H,K) \rightarrow (\twi H, \twi K)$ where
\label\exponent
\eq \eqalign{
\twi H&={2(d-3)\over(d-2)^2}\,K-{d-4\over d-2}\,H\cr
\twi K&=2\,H+{d-4\over d-2}\,K\ .\cr}
\eeq
Hence the effect of the
duality transformation of the Einstein frame solutions is
more complicated in higher dimensions, and one does not find that
the duality simply exchanges the mass and the dilaton charge except
for $d=4$. The standard Schwarzschild geometry in isotropic
corresponds to $(H,K)=(1,0)$. This
nonsingular black hole is mapped to
the dual solution $(\Hbar,\Kbar) = \left(- \; {d-4\over d-2},
2 \right)$. Thus, since $\Hbar < 0$, the dual counterpart of
the black hole in $d > 4$ dimensions always has a negative mass. It is
only for $d=4$ that the mass vanishes to leading order in $\ap$,
and so a more detailed treatment
is required to determine its sign. The self-dual solutions
in $d$ dimensions are $(H,K)=({1\over\sqrt{d-2}},\sqrt{d-2})$,
and $\left(-\, {1\over\sqrt{d-2}},-\sqrt{d-2}\right)$.

In the general case for $d\ge4$, one has
\label\asimptop
\eq
\eqalign{
A_t &\simeq 2v + {Q_\sse\over(d-3)r^{d-3}}+\cdots\cr
e^\phi &\simeq 1-{Q_\ssd\over r^{d-3}}+\cdots\cr
g_{tt}&\simeq -1+{2\mu\over(d-2)r^{d-3}}+\cdots\cr
g_{ij}&\simeq\delta_{ij}\left(1+{2\mu\over(d-3)(d-2)r^{d-3}}+\cdots
\right)\cr}
\eeq
where $\mu=8\pi\GN M/A_{d-2}$ which reduces to $\mu=2\GN M$ for
$d=4$. Now it is straightforward to implement a duality transformation.
Notice that, as discussed in the main text, when $v\ne0$ one must
scale the time coordinate and shift the dilaton by a constant in
order to preserve the desired asymptotic limits for $g_{tt}$
and $e^\phi$. One then finds the following physical
charges for the dual solutions,
\label\duocharg
\eq
\eqalign{
\Qbar_\sse&={1\over1-v^2}\left[(1+v^2)Q_\sse+4v{d-3\over d-2}(\mu-
Q_\ssd)\right]\cr
\Qbar_\ssd&={1\over1-v^2}\left[\left({d-4\over d-2}-v^2\right)Q_\ssd
+{2\mu\over d-2}+{v\,Q_\sse\over d-3}\right]\cr
\twi{\mu}&={1\over1-v^2}\left[2{d-3\over d-2}Q_\ssd-
\left({d-4\over d-2}+v^2\right)\mu-v\,Q_\sse\right]\cr}
\eeq
where again we have assumed that $v^2<1$. Note that for $d=4$, these
results reduce to the corresponding formulae given in eq.~\longeq.
Similarly one finds that eq.'s~\simpeq\ and \simpeqq\ 
are replaced
by
\label\xchangt
\eq
\eqalign{
\mu+{v\over2}Q_\sse&\leftrightarrow 2{d-3\over d-2}Q_\ssd
-{d-4\over d-2}\mu-{v\over2}Q_\sse\cr
\left({2({\twi{\mu}}-\Qbar_\ssd)\over d-2}\right)^2-&\left({\Qbar_\sse\over
d-3}\right)^2=\left({2(\mu-Q_\ssd)\over d-2}\right)^2-\left({Q_\sse\over
d-3}\right)^2\cr}
\eeq
for $d\ge4$. Hence the action of the duality transformations on the
physical charges is very similar in all dimensions.

\listrefs

\bye